\documentclass{aa}
\usepackage[varg]{txfonts}
\usepackage{caption}
\usepackage{subcaption}
\usepackage{hyperref}
\usepackage{natbib}
\bibpunct{(}{)}{;}{a}{}{,}

\begin{document}

\title{VLA detects CO(1--0) emission in the $z=3.65$ quasar SDSS J160705+533558}
\subtitle{}
\author{J. Fogasy \inst{1,2}
        \and K. K. Knudsen \inst{1}
        \and E. Varenius \inst{1}}
\institute{Department of Space, Earth and Environment, Chalmers University of Technology, Onsala Space Observatory, 439 92 Onsala, Sweden
\and Konkoly Observatory, ELKH Research Centre for Astronomy and Earth Sciences, Konkoly Thege Mikl\'os \'ut 15-17, H-1121 Budapest, Hungary\\ \email{fogasy.judit@csfk.org}}
\date{Received 17 November 2021; accepted 11 February 2022}
\abstract{We present CO(1--0) observations of the high-redshift quasar SDSS J160705+533558 ($z=3.653$) using the Karl G. Jansky Very Large Array (VLA). We detect CO emission associated with the quasar and at $\sim16.8\,\rm kpc$ projected distance from it, separated by $\sim800\,\rm km\,s^{-1}$ in velocity. The total molecular gas mass of this system is $\sim5\times10^{10}\,\rm M_{\sun}$.
By comparing our CO detections with previous submillimetre (submm) observations of the source, an offset between the different emission components is revealed: the peak of the submm emission is offset from the quasar and from the CO companion detected in our VLA data. To explain our findings, we propose a scenario similar to that for the Antennae galaxies: SDSS J160705+533558 might be a merger system in which the quasar and the CO companion are the merging galaxies, whose interaction resulted in the formation of a dusty, star-forming overlap region between the galaxies that is dominant at the submm wavelengths.}

\keywords{quasars: individual: SDSS J160705+533558 --
                galaxies: active --
                galaxies: high-redshift --
                galaxies:  --
                galaxies: evolution -- }
\maketitle

\section{Introduction}
\label{sec:intro}
Active galactic nuclei (AGNs) have been considered as the key for understanding the evolution and growth of local massive elliptical galaxies and for explaining the observed tight scaling relations found between the mass of the central supermassive black hole (SMBH) and the mass and velocity dispersion of the bulge component \citep[e.g.][]{1998AJ....115.2285M, 2000ApJ...539L..13G, 2001ApJ...547..140M, 2002ApJ...574..740T, 2003ApJ...589L..21M, 2004ApJ...604L..89H, 2013ApJ...764..184M}. AGNs not only harbour growing SMBHs, but often show signs of intense star-formation observed in infrared and submillimetre (submm) wavelengths \citep[e.g.][]{2001ApJ...555..625C, 2001A&A...374..371O, 2005A&A...440L..51M, 2013ApJ...773...44W, 2016ApJ...816...37V}. In addition, high-resolution observations revealed close, gas-rich companion galaxies in several AGNs, which supports the idea that galaxy mergers play an important role in the build-up of local massive galaxies \citep[e.g.][]{2012ApJ...752L..30W, 2013ApJ...763..120C, 2015ApJ...806L..25S, 2017ApJ...836....8T, 2017MNRAS.465.4390B, 2018MNRAS.479.1154B, 2017A&A...605A.105C, 2017Natur.545..457D, 2018ApJ...856L...5F}.

In this paper we focus on the broad absorption line quasar SDSS J160705-533558 ($z=3.653$, \citealt{2006ApJS..165....1T}).
Based on observations with the Infrared Array Camera (IRAC) and the Multiband Imaging Photometer (MIPS) on board the \textit{Spitzer} Space Telescope, the quasar is associated with a strong rest-frame mid-infrared emission \citep{2004ApJS..154...54L, 2005AJ....129.1198H}. Follow-up submm observations with SCUBA mounted on the James Clerk Maxwell Telescope and the Submillimeter Array (SMA) detected high submm fluxes at $850\,\mu\rm m$ \citep{2009ApJ...698L.188C}.
With a total infrared luminosity of $\sim10^{14}\,\rm L_{\odot}$, SDSS J160705+533558 is one of the most luminous galaxies in the universe.\ These galaxies are commonly referred to as hyper-luminous infrared galaxies (HLIRGs).

An interesting feature of this source that was revealed by SMA observations is that the submm emission is extended over $5.2\arcsec$ ($36\rm\,kpc$) and consists of a clump to the north and a tail towards south. The peak of the $850\,\mu\rm m$ emission is offset from the position of the quasar by $\sim$1.5$\arcsec$ ($\sim$10\,kpc) to the north. \citet{2009ApJ...698L.188C} argued that this separation and the extended nature of the submm emission, supplemented by the \textit{Spitzer} data, indicate that two distinct sources account for the observed emission: an AGN component that dominates the rest-frame mid-infrared and an obscured starburst component with a star formation rate of  $\sim$1000$\,\rm M_{\sun}\,yr^{-1}$ that causes the bright submm emission. 
Moreover, the submm tail might be a tidal structure resulting from the interaction of these two components, which might also trigger the quasar activity and the star formation in this system.

The next step in uncovering the nature of this source is to study its molecular gas content. With this aim, we obtained CO(1--0) line observations of the quasar using the Karl G. Jansky Very Large Array (VLA). In section \ref{sec:observations} we present the details of the VLA observations, in section \ref{sec:analysis} we present the results and describe the data analysis, in section \ref{sec:discussion} we discuss our findings, and in section \ref{sec:conclusion} we summarise our results. Throughout the paper we assume a  $\Lambda$CDM cosmology with $H_{0}=67.2\ \rm{km \ s^{-1} Mpc^{-1}}$, $\rm \Omega_{m}=0.315,$ and $\rm \Omega_{\Lambda}=0.685$ \citep{2014A&A...571A..16P}.

\section{Observations}
\label{sec:observations}
SDSS J160705+533558 was observed with the VLA between 18-27 April 2017 in D configuration and K band (project code 17A-003; PI: Fogasy). The on-source integration time was 68.7 min. The total bandwidth of the observations is 1024 MHz, covering eight spectral windows with 128 channels in each. The data calibration was done applying the VLA Common Astronomy Software Applications (\textsc{casa}\footnote{{\tt https://casa.nrao.edu}}, \citealt{2007ASPC..376..127M}) Calibration Pipeline (v.5.6.1-8), which includes standard calibration and reduction steps such as flagging, bandpass calibration, and flux and complex gain calibration. The quasar J1549+5038 was used as the phase calibrator source, and the flux calibration was done using the standard flux reference source 3C286.
We assumed a 20\% absolute flux calibration uncertainty, which is not included in the flux densities reported in this paper.

The imaging of data was carried out with \textsc{casa} using the \textsc{tclean} algorithm and Briggs weighting with a robust parameter of 0.5. The resulting continuum image has an rms of 11$\, \mu\rm Jy\,beam^{-1}$ and a beam size of $3.25\arcsec\times2.32\arcsec$ with a P.A. of $77.2\degr$. The continuum
was fitted using the line-free channels of every spectral window and was subtracted from the spectral line data in the $uv$-plane using the \textsc{uvcontsub} task in \textsc{casa}. The line data were imaged with a channel width of 6, resulting in a velocity binning of $72.6\,\rm km\,s^{-1}$, and they have an rms of $90\,\rm\mu Jy\,beam^{-1}$. The median restoring beam of the line data is $3.26\arcsec\times2.35\arcsec$ with a P.A. of $76\degr$.

\section{Results and analysis}
\label{sec:analysis}
\subsection{Continuum emission}
\begin{figure}
        \centering
        \includegraphics[width=\columnwidth]{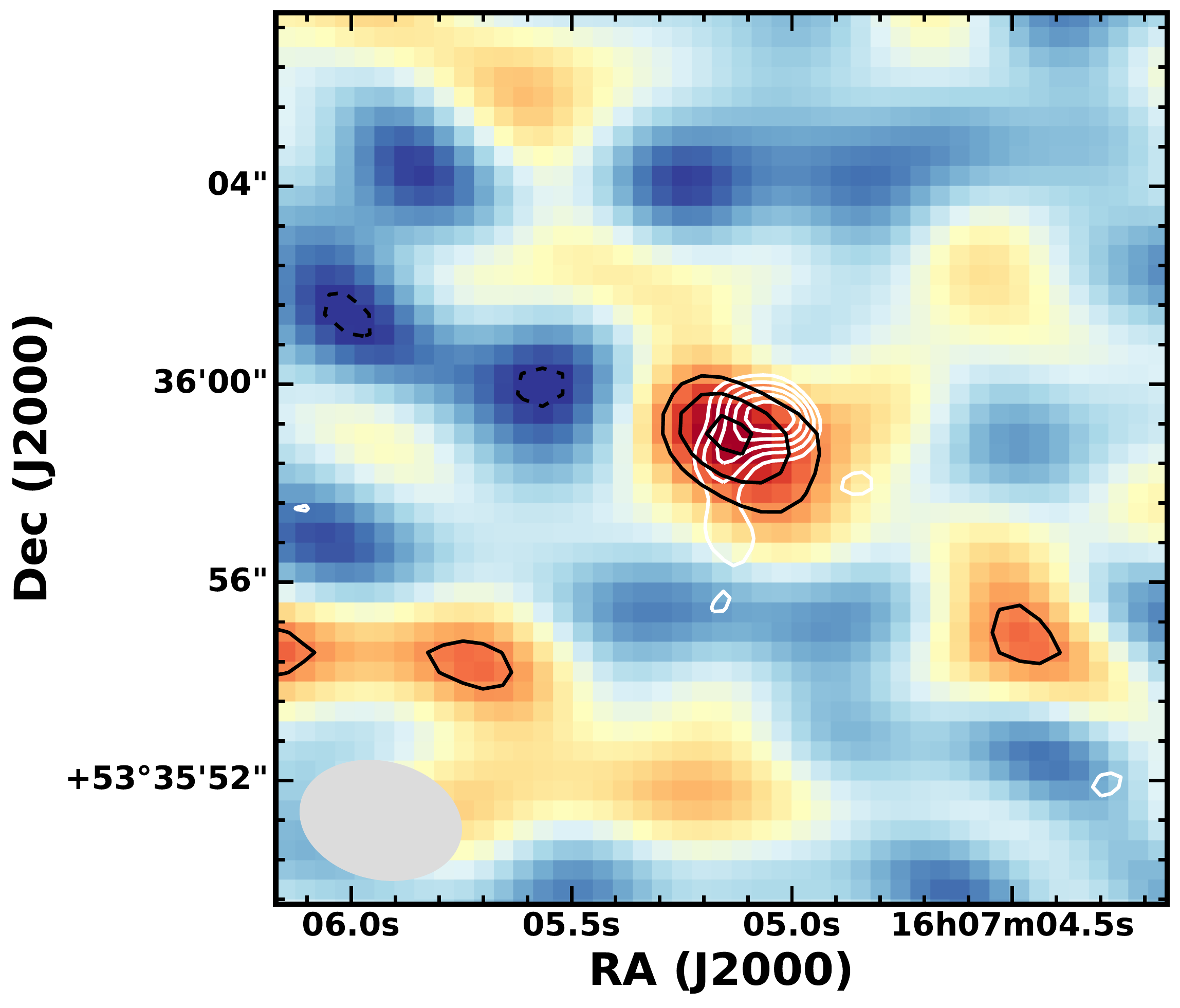}
        \caption{VLA 29.97 GHz continuum emission of the quasar SDSS J160705+533558. The black contour levels are at $[-3,3,4,5]\times\sigma$ ($\sigma=11\,\mu\rm Jy$). Negative contours are shown as dashed lines. White contours show the $850\,\mu\rm m$ SMA continuum from \citet{2009ApJ...698L.188C} with contour levels at $[3,4,5,6,7]\times\sigma$ ($\sigma=1.2\,\rm mJy$). The VLA beam is shown as a grey ellipse in the bottom left corner.}
        \label{fig:sdss_cont}
\end{figure}

We detect VLA 29.97 GHz continuum emission at the position of the quasar, with an integrated flux density of $63.7\pm4.7\,\mu\rm Jy$ (Figure \ref{fig:sdss_cont}). The resolution of the VLA data does not allow us to resolve the continuum emission.  However, the continuum peak is clearly centred on the quasar and not on the offset submm emitting component reported in \citet{2009ApJ...698L.188C}. As the continuum emission is likely synchrotron emission, the submm tail detected by the SMA might be contaminated by synchrotron emission.
We do not detect any other continuum sources in the field.

\subsection{CO(1--0) line emission}
\begin{figure*}
        \centering
        \includegraphics[width=1.0\columnwidth]{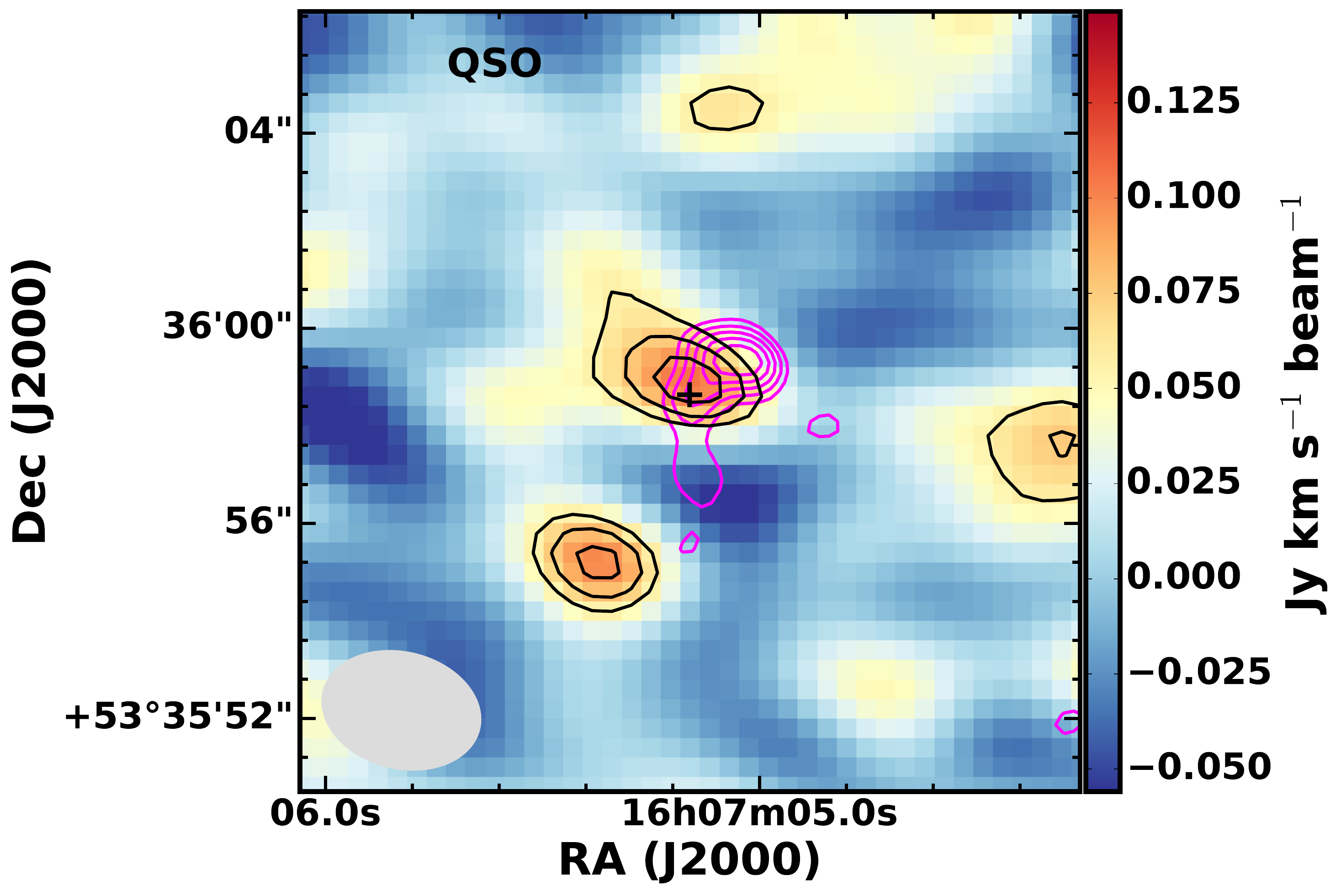}
        \includegraphics[width=1.0\columnwidth]{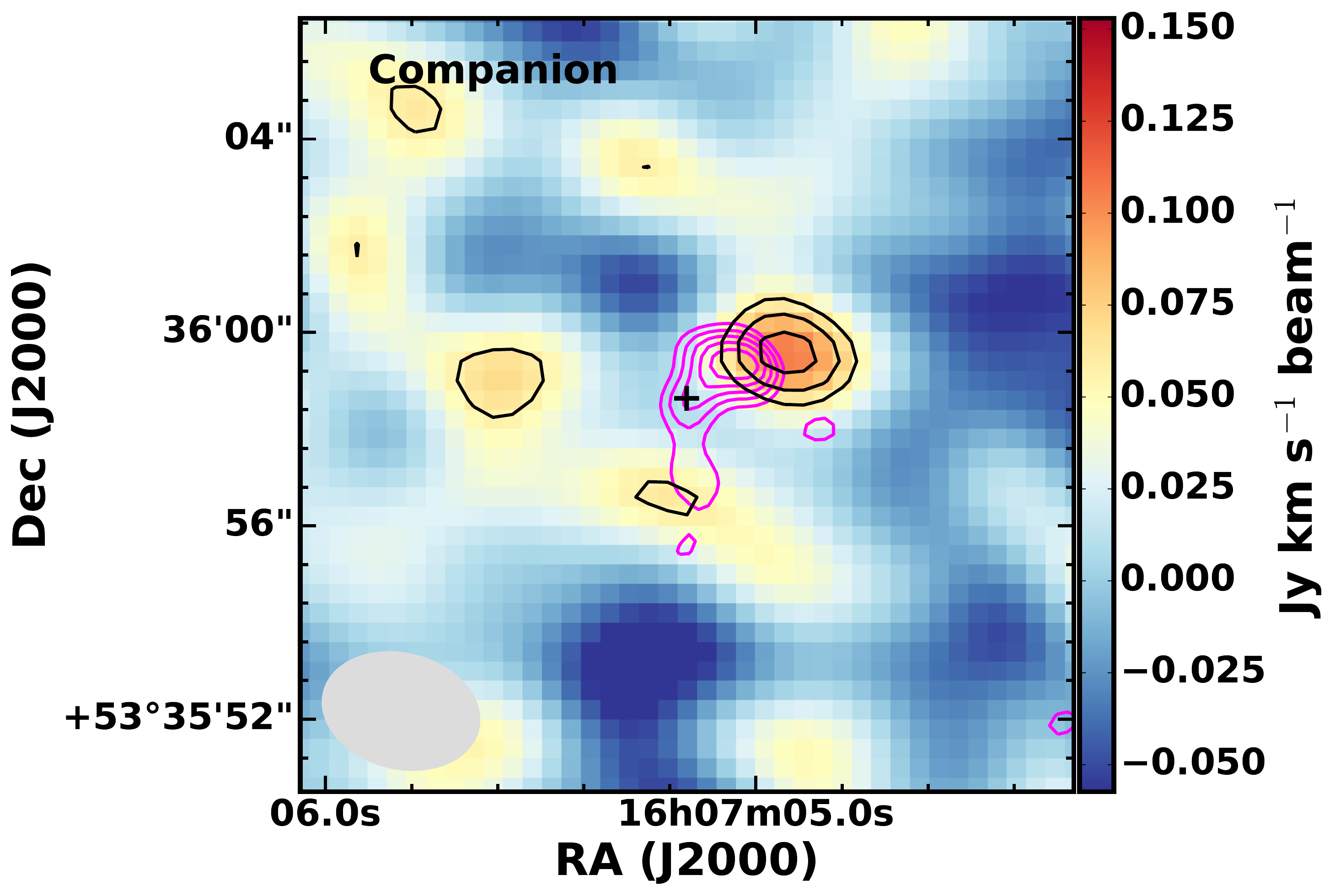}
        \includegraphics[width=\columnwidth]{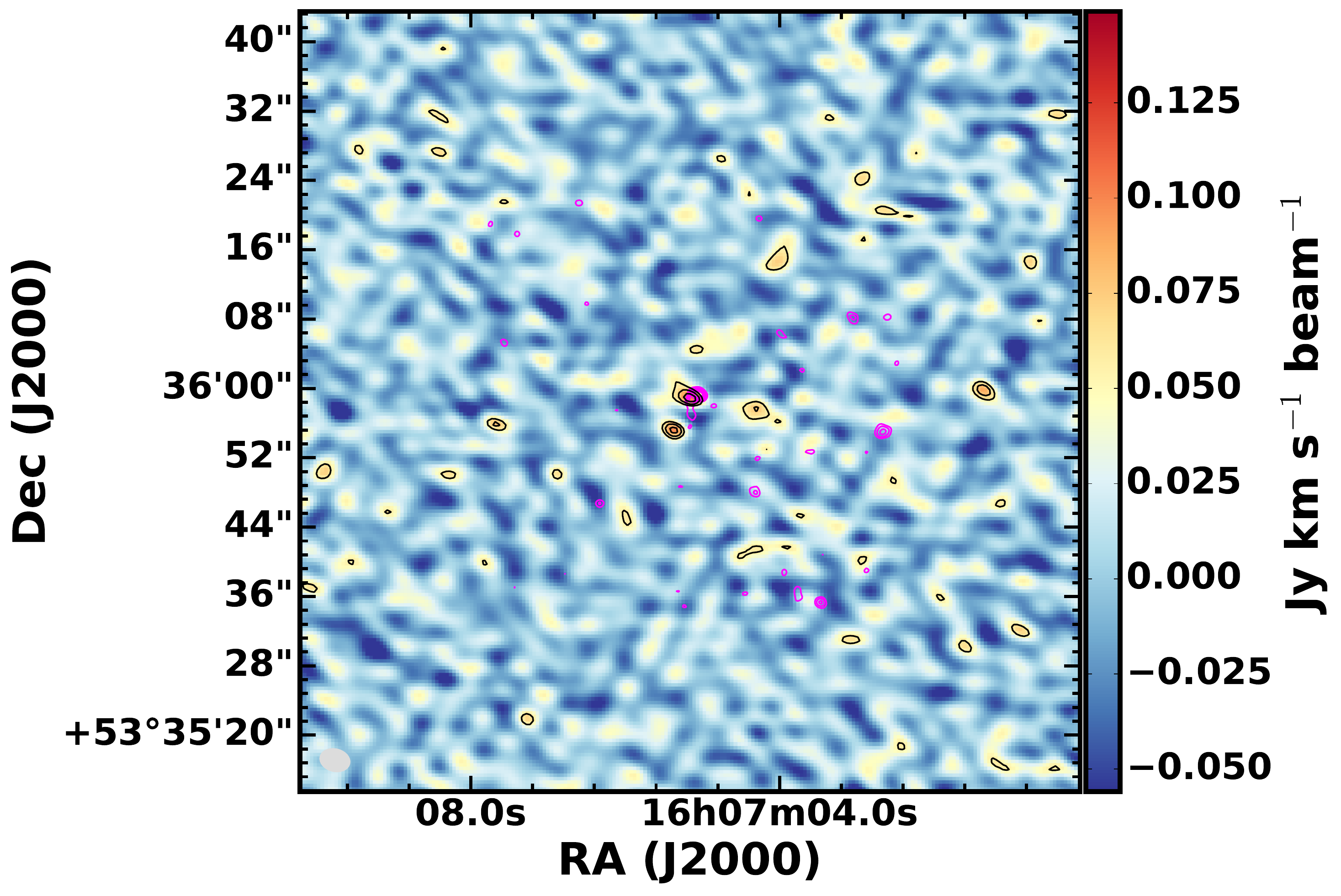}
        \includegraphics[width=\columnwidth]{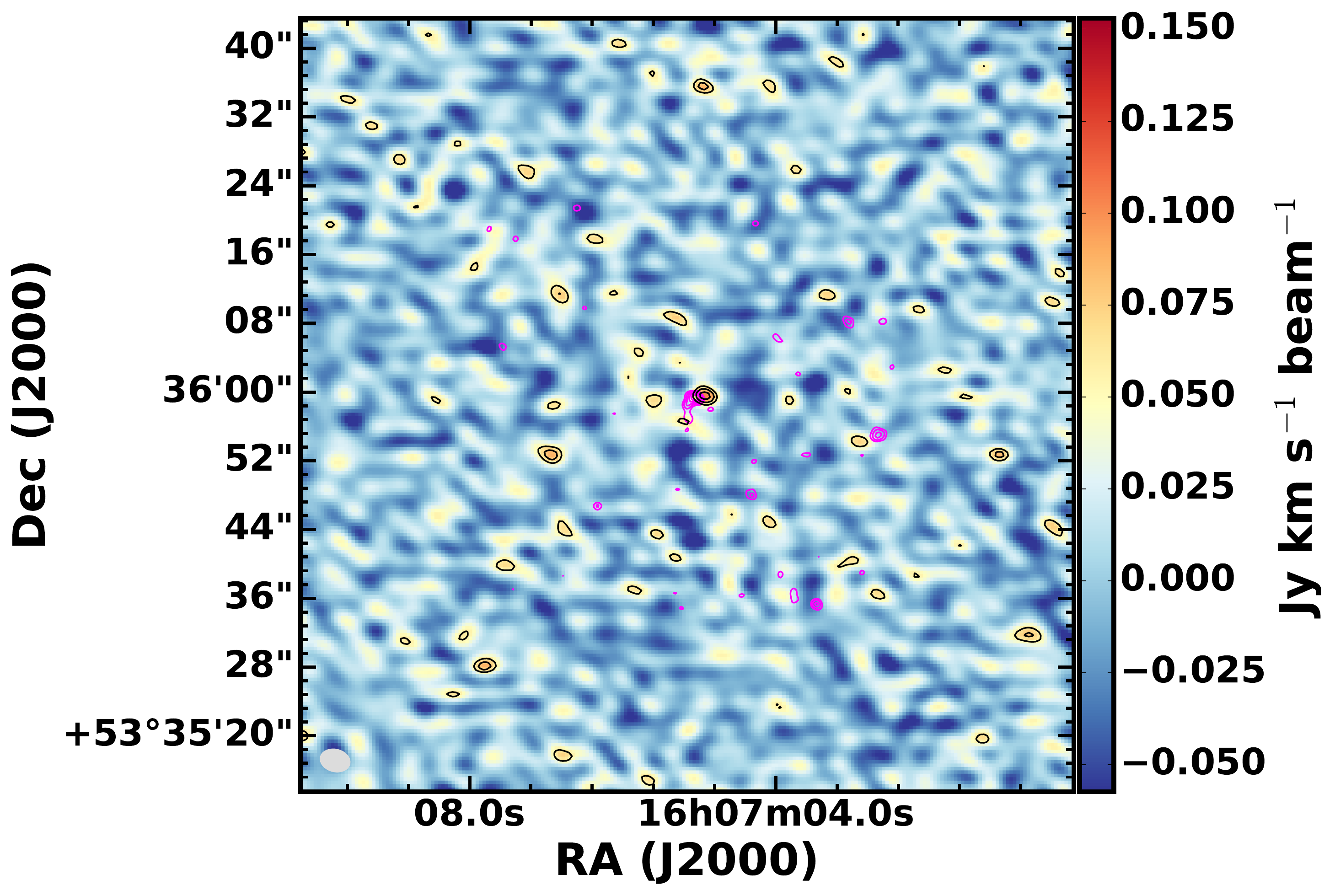}
        \caption{CO(1--0) emission of SDSS J160705+533558. \textit{Top}: Integrated intensity (moment-0) maps of the quasar and the companion. The moment-0 maps are integrated over the velocity range of [-267,382] $\rm km\,s^{-1}$ ([24.74, 24.79] GHz) in the case of the quasar and [-1288, -426] $\rm km\,s^{-1}$ ([24.801, 24.88] GHz) for the companion. The black contour levels are at [3, 4, 5] $\times\,\sigma$, where $\sigma=0.019\,\rm Jy\,beam^{-1}\,km\,s^{-1}$. The magenta contours show the $850\,\mu\rm m$ SMA continuum of the source, with the same contour levels as in Fig. \ref{fig:sdss_cont}. The black cross marks the position of the quasar. The beam size is indicated as a grey ellipse in the bottom left corner. \textit{Bottom}: Large-scale moment-0 maps of the quasar and the companion.}
\label{fig:FOV_m0}
\end{figure*}

\begin{figure*}
        \centering
        \includegraphics[width=0.67\columnwidth]{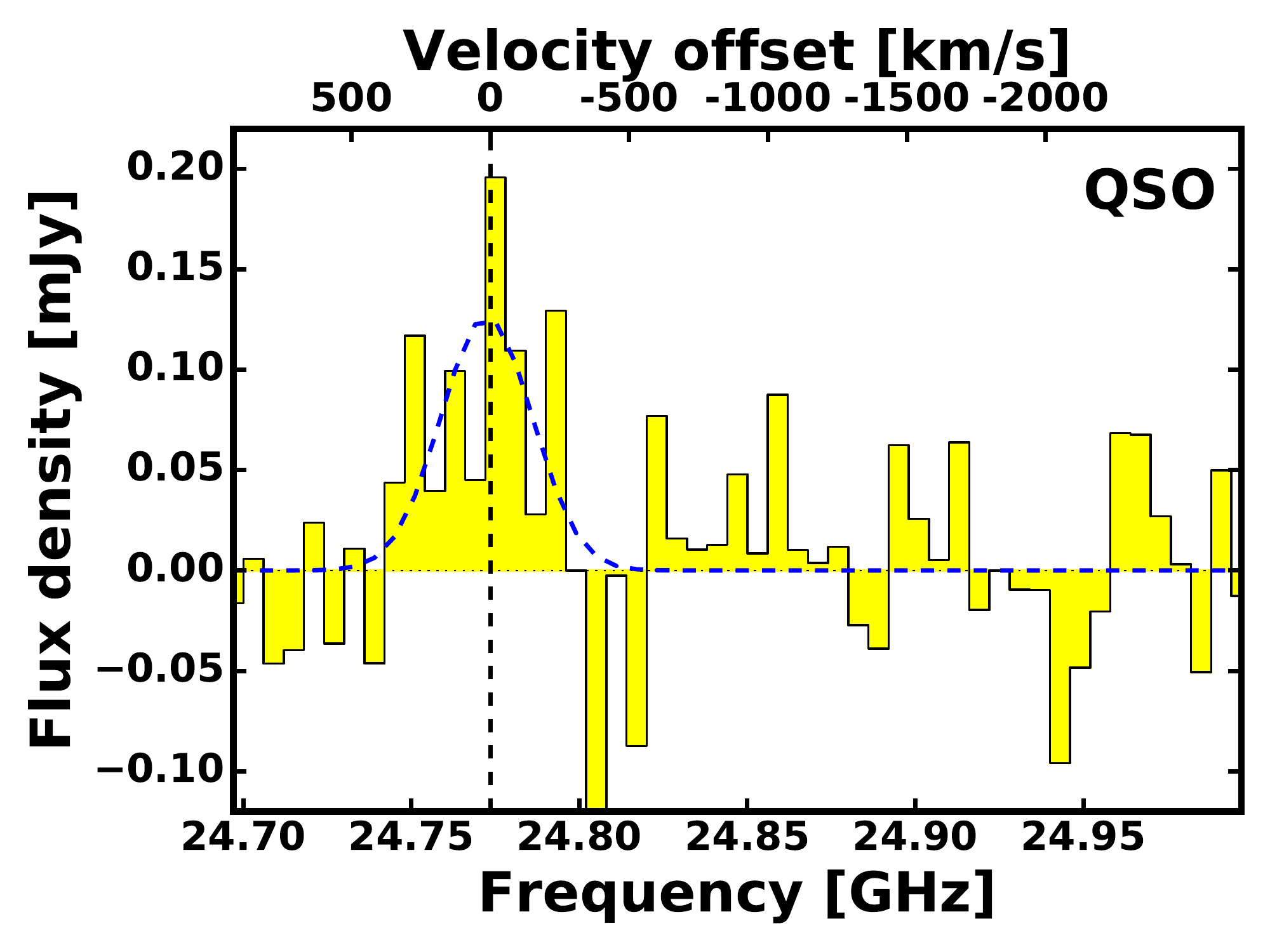}
        \includegraphics[width=0.67\columnwidth]{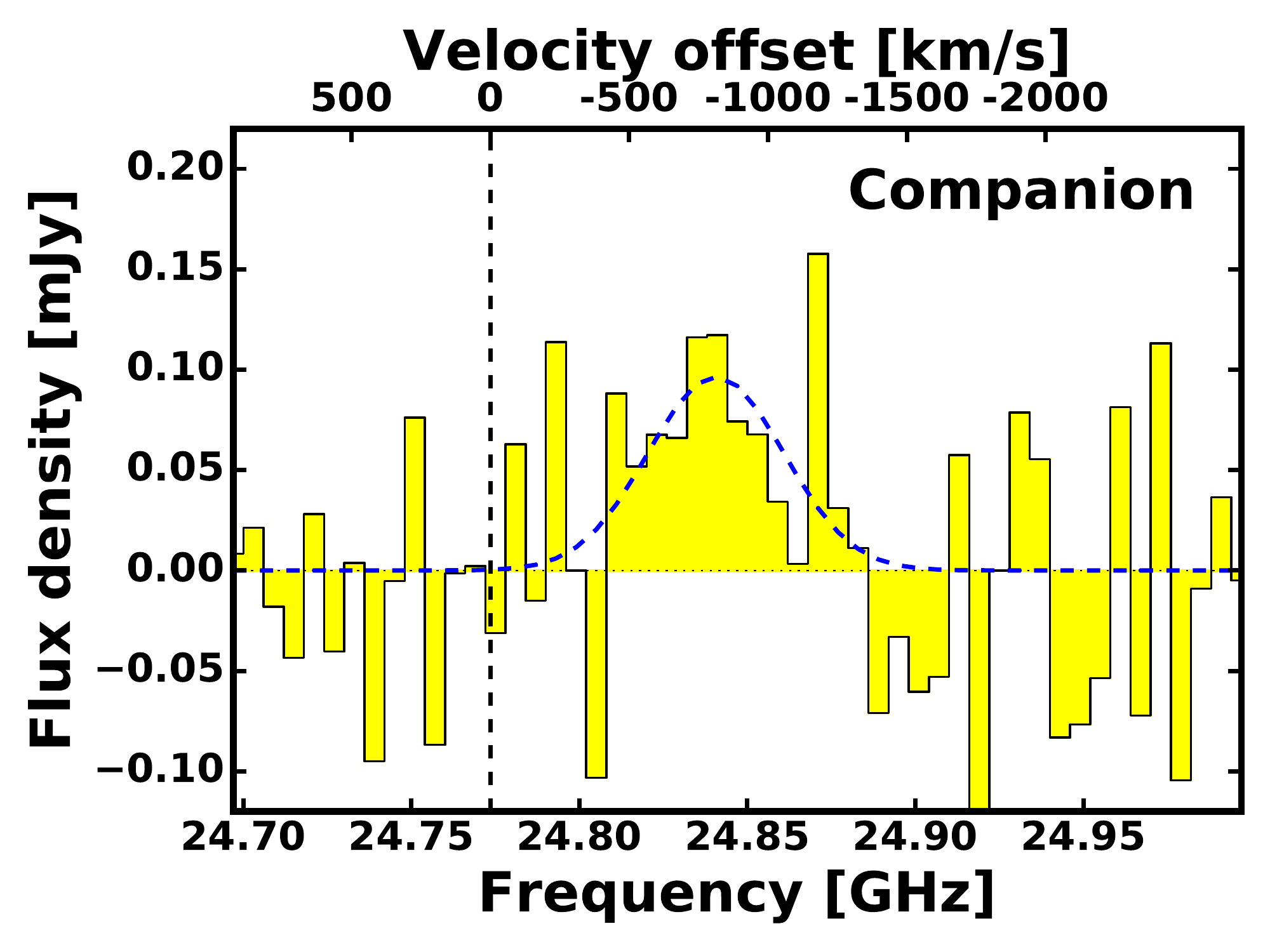}
        \includegraphics[width=0.67\columnwidth]{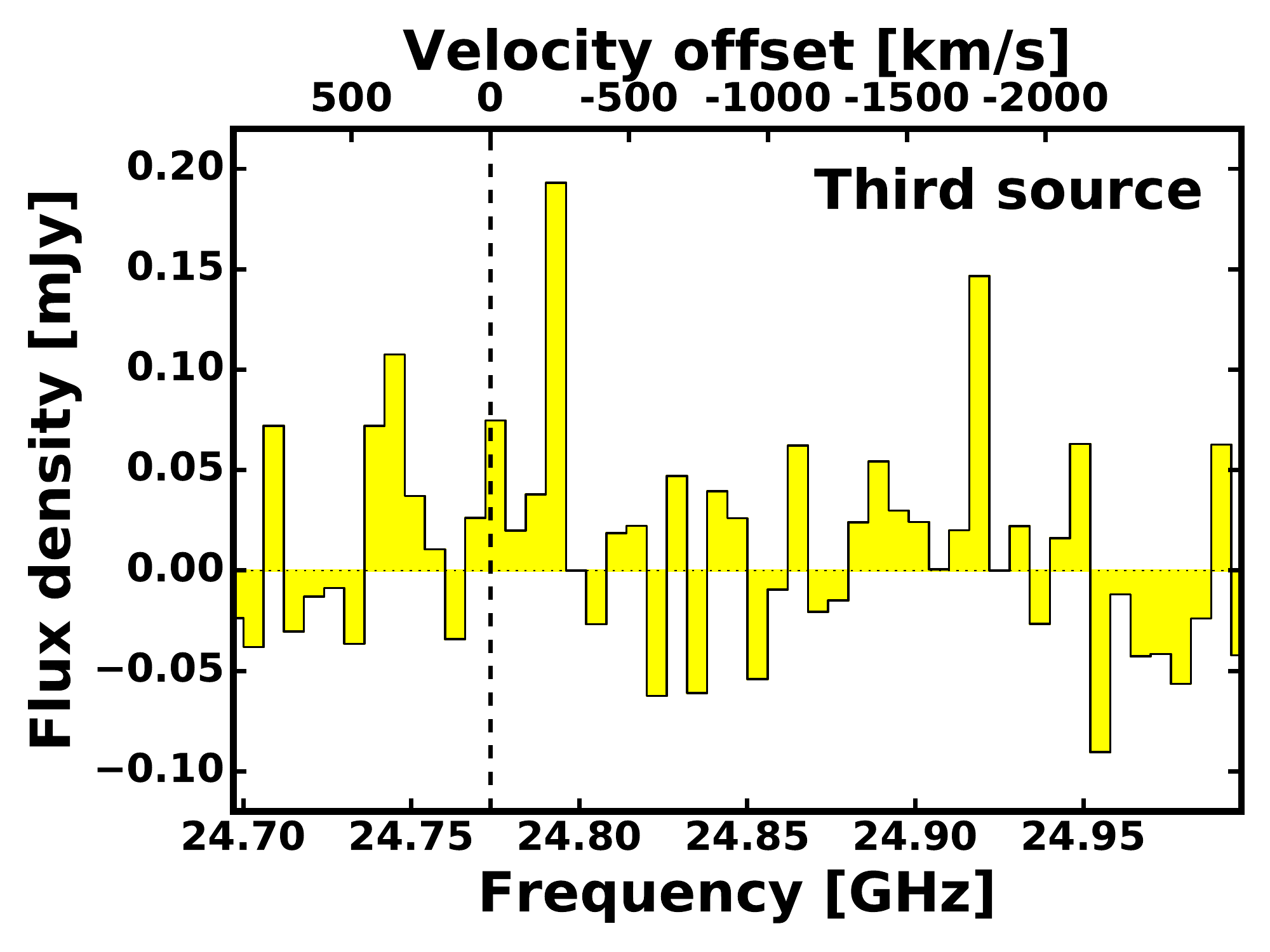}
        \caption{CO(1--0) emission line of SDSS J160705+533558. The extracted CO(1--0) line spectra of the quasar, its companion, and the additional source to the south of the quasar. The spectra are binned to $72.6\,\rm km\,s^{-1}$ per channel, and the continuum is subtracted. The dashed blue curves show the Gaussian fits to the line profiles. The vertical dashed line indicates the optical redshift of the quasar ($z=3.653$). The top-axis show the relative velocity offset with respect to the optical redshift of the quasar.}
\label{fig:line}
\end{figure*}
We detect CO(1--0) line emission from three regions, one at the position of the quasar, one at $2.3\arcsec$ ($\sim16.8\,\rm kpc$) distance from it (hereafter referred to as companion), and another source to the south at $3.9\arcsec$ ($\sim28.5\,\rm kpc$) distance from the quasar (Figure \ref{fig:FOV_m0} top panels). Based on the velocity-integrated intensity (moment-0) maps of the sources, the line-emitting regions are compact, and all three sources are detected at $5\sigma$ level. The rms of the moment-0 maps is $\sigma=0.019\,\rm Jy\,beam^{-1}\,km\,s^{-1}$ and is measured from the emission-free regions of the images. While there are several detections at $4\sigma$ level in the larger field-of-view moment-0 maps (Figure \ref{fig:FOV_m0} bottom panels), based on \citet{2012ApJ...753..135K}, we set a $5\sigma$ detection limit in order to exclude spurious detections from the analysis. Another way to exclude spurious detections is to only consider sources with counterparts in other observing bands \citep[e.g.][]{2012MNRAS.426..258A}. In the case of the quasar, this criterion is fulfilled as the centre of the CO emitting region coincides with the optical position of the quasar. Because the CO companion is very close to the peak of the SMA detection of SDSS J160705+533558, we also consider it to be real. In contrast, the additional source to the south has neither a counterpart nor a clear line profile (Figure \ref{fig:line} right panel), and it is most likely a spurious detection. Thus, we do not include it in the further analysis.

To determine the line properties of the sources, we extracted their spectra from the line data cube. For this we used the moment-0 maps and made a $3\sigma$ cut, considering emission only from this region, then we used a small subcube around the sources to extract their spectra. We fitted the spectra with single Gaussian line profiles (Figure \ref{fig:line}). Based on the fit results, the velocity-integrated flux densities of the quasar and the companion are $0.052\pm0.020\,\rm Jy\,km\,s^{-1}$ and $0.058\pm0.028\,\rm Jy\,km\,s^{-1}$, respectively, where the errors reflect the uncertainty of the Gaussian fit (Table \ref{tab:table1}). As an alternative to the Gaussian fit results, a simple integration over the line channels yields a velocity-integrated flux density of $0.058\pm0.011$\,Jy\,km\,s$^{-1}$ for the quasar and $0.064\pm0.015$\,Jy\,km\,s$^{-1}$for the companion, where the error is based on error propagation of the rms of the line-free channels. These results are consistent with the Gaussian fit results, although the error is slightly lower. For further analysis, we used the results of the Gaussian fits.
The CO(1--0) line profile of the quasar is narrow, with a full width at half maximum (FWHM) of  $387\pm114\,\rm km\,s^{-1}$ , while the companion has a broader line profile with $\rm FWHM=572\pm208\,\rm km\,s^{-1}$. These line widths are consistent with the results found for other high-$z$ quasars and starbursts \citep[e.g.][]{2013ARA&A..51..105C}. The CO and optical redshifts of the quasar are consistent within the uncertainties, but the CO line of the companion is shifted by $-800\,\rm km\,s^{-1}$ with respect to the optical redshift of the quasar. 

We derived the CO line luminosities of the sources using the following equation: $L_{\rm CO}^{'}=3.25\times10^{7}\times S_{\rm CO}\Delta V\times\nu_{\rm obs}^{-2}\times D_{L}^{2}\times(1+z)^{-3}$, where $S_{\rm CO}\Delta V$ is the velocity-integrated flux, $\nu_{\rm obs}$ is the observed frequency of the line in GHz, and $D_{L}$ is the luminosity distance in Mpc \citep{1997ApJ...478..144S}.
The CO line luminosities of the quasar and the companion are $(3.0\pm1.1)\times10^{10}\,\rm K\,km\,s^{-1}\,pc^{2}$ and $(3.3\pm1.6)\times10^{10}\,\rm K\,km\,s^{-1}\,pc^{2}$, respectively. To estimate the molecular gas mass of the sources, we assumed a $\rm CO-H_{2}$ conversion factor of $0.8\,\rm M_{\sun}(K\,km\,s^{-1}\,pc^{2})^{-1}$, which is typically used for starburst galaxies \citep{2005ARA&A..43..677S,2013ARA&A..51..105C}. This yields a molecular gas mass of $(2.4\pm0.9)\times10^{10}\,\rm M_{\sun}$ for the quasar host galaxy and $(2.6\pm1.3)\times10^{10}\,\rm M_{\sun}$ for the companion.

As the peak of the SMA submm continuum is offset from the quasar, we compared the positions of the CO companion and the offset submm emission. In the top right panel of Figure \ref{fig:FOV_m0} we mark the position of the quasar and overlay the $850\,\mu\rm m$ SMA continuum and the moment-0 contours of the companion. By comparing the CO and submm data, there is a $\sim1.4\arcsec$ ($\sim10.2\,\rm kpc$) offset between their peak. This offset might be related to the resolution of the VLA data and all the factors that affect the positional accuracy of interferometric measurements. To estimate the systematic offset of our observations due to thermal noise, we used Eq. (1) from \citet{1988ApJ...330..809R}: given the signal-to-noise ratio of the sources ($\sim5$) and the beam size ($\sim3\arcsec$), the position measurement uncertainty of the VLA data is about $0.3\arcsec$, which is below the measured offset between the CO and submm emission peak. In addition, there may be offsets in the absolute astrometry of the observations due to imperfect phase-calibration. However, the good agreement between the peak of the VLA quasar emission and the tail of the SMA observations suggest that the absolute astrometry, and with that the relative astrometry, is good enough to confidently distinguish the aforementioned three positions. This suggests that this system consists of three components: a submm-faint but gas-rich quasar, a submm-bright, dusty starburst component, and a gas-rich component.

\begin{table}
        \centering
        \caption{Positions, continuum fluxes, and line properties of SDSS J160705+533558.}
        \label{tab:table1}
        \resizebox{\columnwidth}{!}{%
        \begin{tabular}{lcc}
                \hline
                \hline
                 & QSO & Companion\\
                \hline
                RA & 16:07:05.17 & 16:07:04.92\\
                Dec. & +53:35:58.9 & +53:35:59.6\\
                 $S_{29.97\,\rm GHz}\, (\mu\rm Jy$)  & $63.7\pm4.7$ & -- \\
                $S\Delta V_{\rm CO(1-0)}\,\rm (Jy\,km\,s^{-1})$ & $0.052\pm0.020$ &  $0.058\pm0.028$\\
                FWHM $\rm\,(km\,s^{-1})$ & $387\pm114$ & $572\pm208$\\
                $z_{\rm CO(1-0)}$ &  $3.6532\pm0.0008$ & $3.6404\pm0.0014$\\
                 $L_{\rm CO(1-0)}^{'}\, (10^{10}\,\rm K\,km\,s^{-1}\,pc^{2})$ & $3.0\pm1.1$ & $3.3\pm1.6$\\
                $M_{\rm H_{2}}\, \rm (10^{10}\,M_{\odot})$ & $2.4\pm0.9$ & $2.6\pm1.3$\\
                \hline
        \end{tabular}}
\flushleft
\end{table}

\section{Discussion}
\label{sec:discussion}
Our VLA CO(1--0) observations reveal a significant molecular gas reservoir associated with the host galaxy of the quasar, with an estimated molecular gas mass of $\sim2.4\times10^{10}\,\rm M_{\sun}$. The line luminosity of the CO(1--0) emission is comparable to that of other high-$z$ quasars found in the literature \citep[e.g.][]{2016ApJ...827...18S,2018MNRAS.479.1154B}. 
Given the offset between the quasar position and the peak of the submm emission, complemented by the submm tail of the source, \citet{2009ApJ...698L.188C} proposed a merger scenario between a dust- and gas-poor quasar and a submm bright, gas-rich galaxy. However, based on our CO(1--0) observations, it is clear that the host galaxy of the quasar is not gas poor. A similar scenario has been played out in the case of other high-$z$ AGNs, where initial submm and low-$J$ CO observations did not detect significant continuum and gas emission, suggesting that these sources are dust and gas poor (e.g. SMM J04135+10277, TXS 0828+193; \citealt{2020MNRAS.493.3744F, 2021MNRAS.501.5973F}). However, follow-up observations with better resolution and sensitivity and tracing higher-$J$ CO transitions revealed the opposite. A possible higher temperature or different excitation properties of the gas in the AGN can shift the peak of the spectral energy distribution to higher frequencies, making it difficult to detect in traditional submm surveys. This leads to a bias in the interpretation, similar to what happens when interpreting non-detections of low-$J$ CO lines in AGN, where the extremely high excitation conditions of the molecular gas favour the detection of high-$J$ transitions \citep{2017MNRAS.468.4205K, 2020MNRAS.493.3744F}.

In addition, we detect CO(1--0) emission at a $\sim16.8\,\rm kpc$ projected distance from the quasar, separated by $\sim800\,\rm km\,s^{-1}$ in velocity, and the peak of the CO emission if offset from that of the submm emission by $\sim10.2\,\rm kpc$. This offset could be explained by the lower resolution of the CO(1--0) data compared to the submm observations, or it could arise from observational errors affecting the astrometric accuracy of the VLA data. As the latter is estimated to be much lower than the offset between the different components ($\sim0.3\arcsec$) and the good agreement of the absolute positions of the quasar and the submm emission, we  propose an alternative scenario for the origin of the observed offset. 

SDSS J160705+533558 might be in a merger event between the gas-rich quasar and the CO companion, with the bright submm emission tracing a dusty, star-forming region between the two merging galaxies. Moreover, the $\sim800\,\rm km\,s^{-1}$ velocity offset between the CO emission of the quasar and the companion and the simultaneous presence of a star-forming region enclosed by these two components indicate that this system already had its first pericentric passage. This scenario is supported by hydrodynamic simulations of galaxy mergers using detailed, explicit models for stellar feedback \citep[e.g.][]{2013MNRAS.430.1901H}. Based on these simulations, strong torques at the first passage lead to shocks, dissipation, and inflow of gas, which fuels rapid star formation. After the first passage starburst and before the final coalescence, star formation can also be triggered in tidal arms and in the bridge regions, where the gas shocks between the two merging galaxies.

This scenario has been found at low redshift in the Antennae galaxies, also known as Arp 244. This source is a merger system consisting of two spiral galaxies that are close to their second encounter, when their discs will start to merge into a single system \citep{2008AN....329.1042K}. Multi-wavelength observations of Arp 244 revealed that the inner region of this merger has three main components: the nuclei of the merging galaxies, which have significant CO(1--0) emission, and a dusty, gas-rich overlap region, which is the most luminous component in the mid-infrared, and is associated with the highest star formation activity and dust temperature \citep[e.g.][]{1995AJ....109..960W,1996A&A...315L..93V,2000ApJ...542..120W,2010A&A...518L..44K}. The overlap region is likely due to the interaction of the merging galaxies, where the gravitational forces dragged and mixed the interstellar medium of the galaxies and triggered vigorous star formation \citep{2010A&A...518L..44K}. This case is very similar to what we see in SDSS J160705+533558, with the addition of an AGN in one of the merging galaxies.

At high-redshift, a similar system to SDSS J160705+533558 is SMM J02399-0136 ($z=2.8$), the first detected and spectroscopically confirmed submm galaxy \citep{1997ApJ...490L...5S, 1998MNRAS.298..583I}.
It consists of four distinct photometric components that might be two or more galaxies, each within $\sim10\,\rm kpc$ distance to each other. The two main components are a BAL quasar (L1) and an obscured, dusty starburst (L2SW) that dominates the submm/mm emission \citep{2010MNRAS.404..198I}, just as in SDSS J160705+533558. 
The majority of the CO emission is coincident with L2SW, although the quasar slightly contributes to it \citep{2010MNRAS.404..198I, 2018ApJ...860...87F}. 
In addition, in the two other UV systems with modest stellar masses, L1N and L2, no CO emission has been detected. \citet{2015ApJ...806..260F} suggested that in this system we witness a merger event between the quasar and L2, with L2SW being the starbursting overlap region, similar to Arp 244.

\noindent

Another interesting aspect of the merger system of SDSS J160705+533558 is the simultaneous presence of an AGN and a starburst component. The early triggering of black hole and star-forming activity during a merger is in contrast with theoretical studies, which suggest that gas-rich major mergers have several phases, with a merger-induced starburst occurring first, followed by the onset of a buried AGN, which becomes optically visible after a blowout phase \citep[e.g.][]{2005Natur.433..604D, 2005MNRAS.361..776S, 2006ApJS..163...50H, 2008ApJS..175..356H}. 
However, in contrast to the above picture, several systematic studies have shown that mergers are not the primary trigger mechanism of AGNs \citep[e.g.][]{2012MNRAS.425L..61S, 2014MNRAS.439..861K, 2017MNRAS.470..755H, 2017MNRAS.466..812V, 2019ApJ...882..141M, 2020MNRAS.tmp.1279M}. Compared to these contrasting views of AGN triggering, the system of SDSS J160705+533558 represents a new scenario. 
A co-existence of an already triggered AGN and starburst like this suggests that gas-rich major mergers are more complex and there is no universal theory to explain the evolution of high-$z$ galaxies including the triggering and growth of their SMBHs and the stellar build-up of the their host galaxies. The best way forward is to observe these high-$z$ systems with high resolution, targeting both dust emission and gas tracers, preferably more than one transition to ensure a less biased interpretation.

\section{Conclusion}
\label{sec:conclusion}
We have presented VLA CO(1--0) observations of the $z=3.653$ quasar SDSS J160705+533558. We detect CO(1--0) emission at the position of the quasar and from a companion at $\sim16.8\,\rm kpc$ projected distance from the quasar, separated by $\sim800\,\rm km\,s^{-1}$ in velocity. The molecular gas mass of the quasar and its CO companion is $(2.4\pm0.9)\times10^{10}\,\rm M_{\sun}$ and $(2.6\pm1.3)\times10^{10}\,\rm M_{\sun}$, respectively, assuming a starburst-like conversion factor of $\alpha_{\rm CO}=0.8\,\rm M_{\sun}\,(K\,km\,s^{-1}\,pc^{2})^{-1}$.
By comparing our CO detections with an $850\,\rm\mu m$ SMA observation of the source, an offset between the different emission components is revealed: the peak of the submm emission is offset from the quasar and from the CO companion detected in the VLA data. In addition to observational errors affecting the astrometric accuracy of the VLA data, a merger scenario similar to the Antennae galaxies could explain the observed offset. SDSS J160705+533558 might be a merger system, with the BAL quasar and the CO companion being the merging galaxies, whose interaction resulted in the formation of a dusty, star-forming overlap region between the galaxies that is dominant at the submm wavelengths.

\begin{acknowledgements}
We thank the anonymous referee for the helpful comments. JF and KK acknowledge support from the Knut and Alice Wallenberg Foundation. The National Radio Astronomy Observatory is a facility of the National Science Foundation operated under cooperative agreement by Associated Universities, Inc.
\end{acknowledgements}
\bibliographystyle{aa}
\bibliography{fogasy_sdss160705}

\end{document}